\documentclass[prl,twocolumn,showpacs,preprintnumbers,amsmath,amssymb,nofootinbib]{revtex4}
\usepackage{subfigure}
\usepackage{graphicx}
\usepackage{bm}

\newcommand{\ee}{\begin{equation}}
 \newcommand{\eee}{\end{equation}}
\newcommand{\ea}{\begin{eqnarray}}
 \newcommand{\eea}{\end{eqnarray}}

\newcommand{\mplank}{M_{\rm p}}

\newcommand{\omegastep}{$\Omega(w_0,\ode)$}

\newcommand{\om}{\Omega_{\rm m}}
\newcommand{\od}{\Omega_{\rm d}}
\newcommand{\ob}{\Omega_{\rm b}}

\newcommand{\osf}{\bar{\Omega}_{\rm sf}}

\newcommand{\ode}{\Omega_{\rm d}^e}

\newcommand{\aeq}{a_{\rm eq}}

\newcommand{\lcdm}{$\Lambda$-CDM}

\preprint{HD-THEP-06-25}

\begin{document}

\author{Michael Doran}
\email{M.Doran@thphys.uni-heidelberg.de}
\affiliation{Institut f\"ur  Theoretische Physik, Philosophenweg 16,
                                                  69120 Heidelberg, Germany}
\author{Georg Robbers}
\email{G.Robbers@thphys.uni-heidelberg.de}
\affiliation{Institut f\"ur  Theoretische Physik, Philosophenweg 16,
                                                  69120 Heidelberg, Germany}
\author{Christof Wetterich}
\email{C.Wetterich@thphys.uni-heidelberg.de}
\affiliation{Institut f\"ur  Theoretische Physik, Philosophenweg 16,
                                                  69120 Heidelberg, Germany}

\title{ Impact of three years of data from the Wilkinson Microwave
    Anisotropy Probe on cosmological models with dynamical dark
    energy}

\begin{abstract}
The first three years of observation of the {\it Wilkinson Microwave
Anisotropy Probe} (WMAP) have provided the most precise data on the
anisotropies of the cosmic microwave background (CMB) to date. We investigate
the impact of these results and their  combination with data from other
astrophysical probes on cosmological models with a dynamical dark
energy component. By considering a wide range of such models, we find that
the constraints on dynamical dark energy are significantly improved compared
to the first year data.
\end{abstract}

\maketitle

\section{Introduction}
Observations of type Ia supernovae 
(SNe Ia) \cite{Astier:2005qq,Riess:2004nr},
structure formation (LSS) \cite{Percival:2001hw,Tegmark:2003ud}
and the cosmic microwave
background (CMB) \cite{Spergel:2006hy,Readhead:2004gy,Goldstein:2002gf} all
agree on an accelerated expansion of our Universe. This rather unexpected
phenomenon can be explained by modifying 4-D
gravity \cite{Dvali:2000hr,Bekenstein:2004ne}
or adding a new component to the total energy momentum tensor. The simplest
such component is a cosmological constant. It fits all current observations
flawlessly and has a simple interpretation in terms of a vacuum energy.
Yet, its observed value
is 120 orders of magnitude off from the naive estimate $\Lambda \sim \mplank^4$,
where $\mplank$ is the reduced Plank mass. The coincidence between
this minute dark energy contribution and the observed energy density of matter 
is rather puzzling. If not given by chance via some sort of anthropic principle,
it necessitates a mechanism that explains this coincidence. An immediate 
possibility is a coupling \cite{Wetterich:1994bg,Amendola:1999er,Amendola:2006qi}
between (dark) matter and dark energy (though there might be problems due 
to quantum effects \cite{Doran:2002bc}). 

Another solution is an attractor behavior 
\cite{Wetterich:fm,Ratra:1987rm,Caldwell:1997ii}
of dark energy that leads to an almost constant ratio between the fractional 
energy density $\od(z)$ of dark energy and the species otherwise dominating the
expansion, i.e. photons and neutrinos during radiation domination and matter
during matter domination. Coincidentally, such an attractor behavior corresponds
to a scalar field with exponential potential that arises in string theories and 
when solving the  cosmological constant problem from the point of 
view of dilatation symmetry \cite{Wetterich:fm}. 
The non-vanishing $\od(z)$ at higher redshifts
alleviates the problem of  explaining the coincidence
of matter and dark energy today $\om^0 \approx \od^0$.
Instead of fine tuning $\Lambda$
to many orders of magnitude, the tuning needed is of the order of $10^{-3}$.
However, the tuning needed
for such \emph{early dark energy} cosmologies increases the less dark energy there
is at earlier times. A detection of early dark energy, on the other hand, would give
crucial hints to fundamental laws of nature. The aim of this study therefore is to
investigate the implications of the three year data of WMAP on dynamical dark energy
models in general and their respective fractions of early dark energy.

In view of the theoretical uncertainties many different techniques
have been employed
in the analysis of the dark energy, ranging from atttempts to reconstruct
the potential of a scalar field dark energy (e.g. Ref. \cite{Sahlen:2006dn}) to the principal
component approach of Ref. \cite{Huterer:2002hy}.  We consider the redshift dependence
of the fractional dark energy $\od(z)$ as a free function to be ``measured''
by observation. We investigate in this note various parameterizations
and an interpolated model. The possible coupling between dark energy
and dark matter is neglected in this study.

\section{Observational tests}
Dark Energy influences the expansion history of our Universe. In particular,
the age $t_0$, conformal horizons of today $\tau_0$ and at last scattering $\tau_{ls}$ 
and the sound horizon $r_s$ at last scattering are modified.
The effects can be understood analytically
 \cite{Doran:2000jt,Doran:2001rw,Bartelmann:2005fc} using  an effective description
in terms of weighted averages relevant for the epoch of last scattering
\ee
\ode \equiv \tau_{ls}^{-1} \int_0^{\tau_{ls}} \od(\tau) 
\eee
and structure formation
\ee
\osf  \equiv [ \ln{a_{\rm tr}}-\ln{\aeq} ]^{-1} \int_{\ln \aeq}^{\ln a_{\rm tr}} \od(a)\  {\rm d} \ln a, 
\eee
where $a_{tr} \approx 1/3$. 
The effect of dark energy on the CMB is twofold. Through the modified expansion
history, it changes the acoustic scale $l_A$. In addition, it leads to a decay of the gravitational
potential that is seen as an integrated Sachs-Wolfe contribution and a suppression of fluctuations on
small scales \cite{Caldwell:2003vp}. 
The suppression of growth can be understood by looking at the equation 
of motion for cold dark matter perturbations inside the horizon, where the dark energy fluctuations
are negligible  \cite{Ferreira:1997hj}:
\ee\label{eqn::deltam}
\ddot \delta_m +  \frac{\dot a}{a} \dot \delta_m - \frac{3}{2} \left(\frac{\dot a}{a}\right)^2 \om \delta_m = 0.
\eee
Here the derivative is with respect to conformal time $\tau$.
In a matter Universe, $\om=1$ and the solution is $\delta_m \propto a$. With dark energy
present, $\om < 1$ and the growth of structure slows down according to the solution of 
\eqref{eqn::deltam}  \cite{Ferreira:1997hj}
\ee
\delta_m \propto a^{[\sqrt{25 - 24\od} -1]/4} \approx a^{1 - 3\od/5}. 
\eee
This suppression starts as soon as a mode enters the horizon. As $\delta_m$ cannot
grow during radiation domination, this leads to a red tilt of the CMB and matter power spectra
up to the scale of the mode entering just at matter-radiation equality $k_{equ}$. 
All modes with $k > k_{equ}$ have been inside the horizon before equality and are suppressed
by the same factor. Hence, early dark energy mimics to some extend a running spectral index with
the important difference that the running stops at $k_{equ}$. All in all, the suppression
leads to a smaller $\sigma_8$ compared to a \lcdm\ universe according to \cite{Doran:2000jt}
\begin{equation} \label{main}
 \frac{\sigma _8 (Q)}{\sigma _8 (\Lambda)}\approx
\left( a _{\rm eq}\right)^{ 3\, \osf / 5}
 \left(1-\Omega _{\Lambda}^0 \right)^{-\left (1+ \bar w ^{-1}\right)/5}
 \sqrt{\frac{\tau _0 (Q)}{\tau _0 (\Lambda)}},
\end{equation} 
where $\tau$ is the conformal horizon today and $\bar w$ is a suitably defined average
equation of state of dark energy. As a rough rule of thumb, an increase of $\osf$ by $10\%$
leads to a decrease of $\sigma_8$ by $50\%$. In the following numerical analysis, we will
not use constraints on the overall normalization of the power spectrum, i.e. we marginalize
over the bias of 2dF and SDSS. As we will see, the data we use nevertheless constrains $\osf$.

In contrast to linear growth, non-linear structure formation is enhanced in early dark energy
cosmologies. The density contrast  $\delta_c$ corresponding to a collapsed structure is lower than in
$\Lambda$-CDM \cite{Bartelmann:2005fc}. 
As the abundance is exponentially sensitive to $\delta_c$, the cluster abundance is considerably
higher for a given $\sigma_8$, as compared to $\Lambda$-CDM.
In particular, the abundance of clusters at 
higher redshift drops more slowly than in $\Lambda$-CDM, which will soon be probed
by gravitational lensing and Sunyaev-Zel'dovich surveys.

\section{Investigated Models}
Constraints on the values of cosmological parameters
are always model dependent.
For this analysis, we therefore select dark energy models with a large variety
of different and in part opposite physical properties. 
This approach allows to identify the model dependencies of the
 best fit ranges for the standard
cosmological parameters, and their sensitivity
 to a change of the underlying dark energy behavior
as compared to the \lcdm\ model.
The first of these dark energy models is a leaping kinetic term model (``LKT''),
where a change of the kinetic term of a scalar field at late times leads to acceleration \cite{Hebecker:2000zb}.
In addition, we consider two models described by a parameterization of
the evolution of the dark energy fraction $\od(a)$. In one of these models \cite{Wetterich:2004pv},
$\od$ rather slowly relaxes from today's $\od^0$ to an asymptotic early-time value of $\Omega_\star$
and $\od$ can be written as $\od = \od(w_0, \Omega_\star)$. The other
parameterization \cite{Doran:2006kp}, which
is  a function of $w_0$ and  $\ode$  exhibits a faster variation of $\od$ and is characterized by a minimum
 amount of dark energy $\ode$ throughout all cosmological epochs. This essentially
fixes  $\od$ to $\ode$ from early times until redshifts of a few.
We also include two models which parameterize the
equation of state $w(a)$ of dark energy. The versatile parameterization
of Bassett et al. \cite{Bassett:2002qu}, generalized
by Corasaniti and Copeland 
in Ref. \cite{Corasaniti:2002vg} (``C\&C''),
has four parameters for the equation of state,
namely its value today $w_0$,
its value during the matter dominated era $w_m$ as well
 as the scale factor $a^m_c$ and width $\Delta_m$ of the transition
between these values, so
that $w = w(w_0, w_m, a^m_c, \Delta_m)$.
In addition, we consider the simple parameterization $w(a) = w_0 + w_1(1-a)$ 
\cite{Chevallier:2000qy,Linder:2002et} frequently used in the literature. 
Finally, we analyze a model with $\od$
linearly interpolated in $\ln a$ between values at 
$z=1,\ 3,\ 10,\ 100$ and $z=1100$, leaving considerable freedom
for the variation of $\od(a)$ at the cost of a rather large number of model parameters.

In addition to their respective dark energy parameters, all models
depend on the standard cosmological parameters: the present matter energy
fraction $\om$ and baryon energy fraction $\ob$, the Hubble
parameter $h$, optical depth $\tau$, scalar spectral index $n_s$ and
the initial scalar amplitude $A_s$, which we took into account using
the observationally relevant combination $\ln(10^{10} A_s) -
2\tau$. We chose flat priors on all parameters.

The equation of state of the LKT and the  $\od(w_0, \Omega_\star)$-models
was not allowed to cross the cosmological constant boundary of $w=-1$, and
their fluctuations were treated like scalar field perturbations.
The other models were allowed to cross $w=-1$. For these models,
the speed of sound $c_s^2 = \delta p / \delta \rho$ was fixed
at $c_s^2=1$, so that the perturbation equations for the pressure fluctuations
remain well-defined even at the crossing.
For the interpolated and the \omegastep\ models, this procedure was
only adopted when the equation of state was close to the crossing,
and the fluctuations were treated as scalar
field perturbations everywhere else, so that these models had
the usual scalar field perturbations during almost the entire evolution.
This treatment is necessary because a single scalar
field cannot traverse $w=-1$
\cite{Vikman:2004dc,Hu:2004kh,Huey:2004jz,Caldwell:2005ai,Zhao:2005vj}.

These models were compared to two different sets of data. Set I are the  WMAP 3-year data,
and set II consists of  WMAP \cite{Spergel:2006hy}, BOOMERANG'03 \cite{MacTavish:2005yk} ,
VSA  \cite{Dickinson:2004yr}, CBI \cite{Readhead:2004gy},
ACBAR \cite{Kuo:2002ua} for the CMB plus 2DF \cite{Percival:2001hw}  and SDSS \cite{Tegmark:2003ud} for LSS
and SNe Ia data \cite{Astier:2005qq,Riess:2004nr} combined. 
We omitted the baryon acoustic oscillation data \cite{Eisenstein:2005su}, as it 
 is currently not as sensitive as CMB and LSS in constraining early dark energy \cite{Doran:2006xp}.

\section{Results}
\begin{figure}
	\centering
	\includegraphics[scale=0.6]{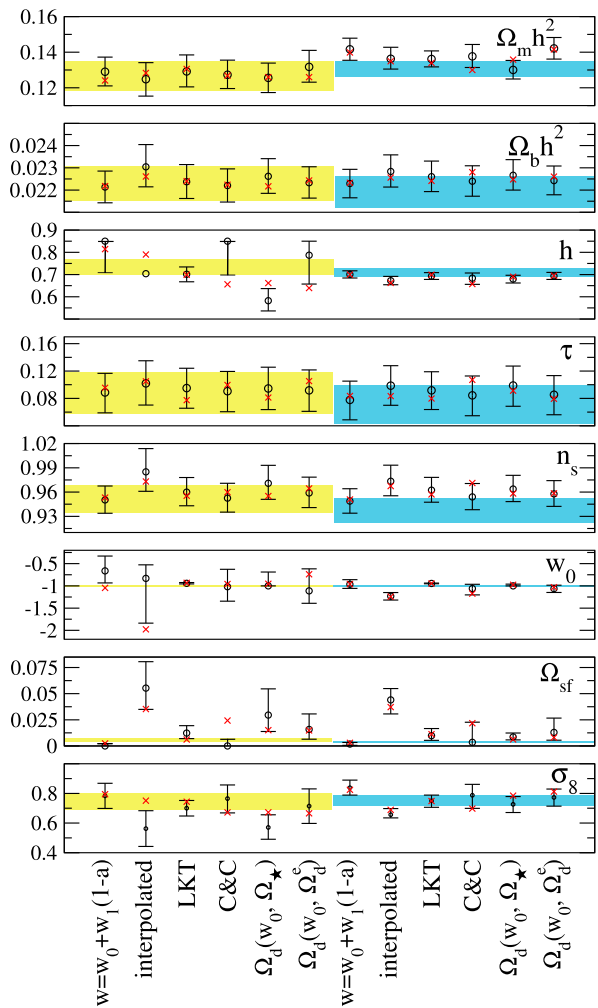}
        \caption{Monte-Carlo results for the cosmological parameters.
                      The left-hand side shows the 1$\sigma$ confidence intervals
                      for the comparison of the models to the 3-year data of WMAP,
                      the right-hand side corresponds to the results for
                      the analysis with the combined set II. The shaded regions depict
                      the 1$\sigma$ intervals for the \lcdm\ analysis from WMAP
                      for the WMAP data only and the ``all" set, respectively. The crosses
                       show the value of the respective parameter for the best fit model
                      in the Monte-Carlo chains.}
        \label{Fig::gross}
\end{figure}
The results are summarized in Figure \ref{Fig::gross}, which yields both the constraints
on each of the standard parameters for all six of the models as well as their scatter. Also shown are the values for
$\osf$ and $\sigma_8$.
As a first result from the Monte-Carlo analysis, we find that cosmological models with a dynamically
evolving dark energy component fit the data as well as \lcdm, but not better. Secondly, the
constraints on the standard cosmological parameters are well compatible with the results found by
WMAP for a \lcdm\ cosmology, almost irrespective of the assumed behavior of the dark energy. Models
that allow for a significant amount of dark energy at early times do have,
however, a few significant features. Most prominently, they have
 a lower $\sigma_8$ from the linear analysis, as expected.
In the light of the rather low scalar spectral index $n_s$ found by WMAP, one 
might have suspected that dynamical dark energy models would allow for a
scalar spectral index that would be closer to $n_s=1$,
which corresponds to a scale invariant
spectrum of the initial fluctuations. However, all investigated dynamical
dark energy models show a preference for $n_s<1$, extending the result found by WMAP to
a wider range of cosmological models.
\begin{figure}
	\centering
	\includegraphics[scale=0.33]{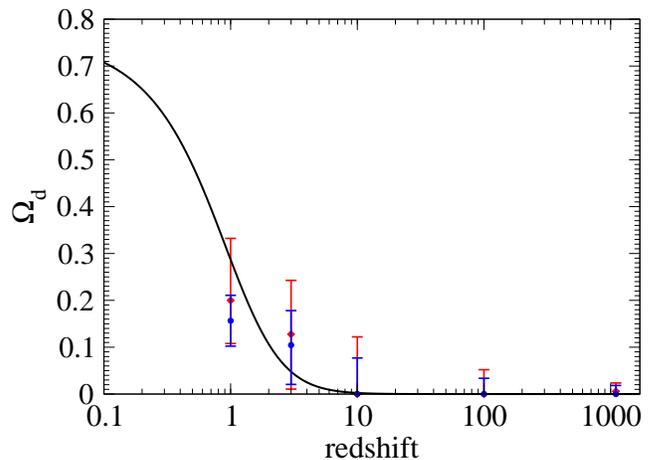}
        \caption{One-sigma confidence intervals for the fraction of dark energy at different
                       redshifts for the interpolated model. Red is the result obtained by comparing
                       to the WMAP 3-year data, blue (lower)  corresponds to the combined set II
                       as described in the text. The black line shows the evolution
                       of $\Omega_\Lambda$ for a \lcdm\ universe
                       with the WMAP 3-year best fit values for the standard cosmological parameters.}
        \label{Fig::omega_v_plot}
\end{figure}

The one sigma bounds on the dark energy fraction for the interpolated model are shown
in  Figure \ref{Fig::omega_v_plot}. It is apparent from this plot that, while
the preferred amount of dark energy in this model includes the values typical for \lcdm\ cosmologies,
it also allows for much larger fractions of dark energy for redshifts $z=3$ and higher.
The evolution of the dark energy density $\sim \od h^2$ can be substantial and is not required to be monotonic in this model.
This can  lead to pronounced ISW contributions. With $\od$ as a free parameter at different $z$, 
the model can to some extend ``manufacture'' the shape of the $TT$-power spectrum at scales
larger than the first peak, leading to a good fit to the WMAP-3 data.  
This, however, comes  at the cost of severely suppressing $\sigma_8$
and is in conflict with the data of set II. Furthermore, with WMAP-3 alone, the Hubble
parameter is considerably less constrained for this model. This is due to the fact that
with $\od(z)$ for $z>1$ given by the model's parameters independent of $h$, the acoustic
scale $l_A$ depends only very weakly on $h$. 
The relative independence of $\od$ in one redshift bin from its neighboring values also
leads  to the Monte-Carlo algorithm finding a comparatively high number
of well-fitting models  with rather high $\osf$.
The interpolated models consequently have the highest values of $\osf$,
with the upper $2\sigma$ limit reaching $\osf \lesssim 11\%$ for WMAP-3 alone
and $6.5\%$ from the combined set II.

The $2\sigma$ upper bounds of the other models for $\osf$ from WMAP-3
alone vary from a very low $0.5\%$, caused by the choice of
parameterization for the \mbox{$w(a) = w_0 + w_1(1-a)$} model, to $\sim 5\%$
for the \omegastep-model and $\sim 7\%$ for the model with $\od
= \od(w_0, \Omega_\star)$. This represents a decrease of about
one to two percent compared to the first year data.  For the combined
set II, the upper bound decreases for all these models to less than
about three percent, with the \omegastep-model yielding the
highest value of $\osf \lesssim 4\% $.
We recall that already a few percent of $\osf$ can have
important effects on the abundance of nonlinear
structure at high redshift \cite{Bartelmann:2005fc}.

\section{Conclusions}
The three year data of WMAP was used to estimate cosmological parameters
for a wide range of dynamical dark energy models. We have shown that this new
data in combination with large scale structure data  constrain the average
amount of dark energy during the time of
structure formation to $\osf \lesssim 4\%$ for a fair sample of  dark energy models 
from the literature. We have also constructed a parameterization
of $\od(a)$ which linearly interpolates between the dark energy
fraction at several redshift bins. This allows for a considerably higher
fraction of $\osf$. The analysis also shows that the values of the
standard cosmological parameters for the dynamical dark energy models
are well compatible with the values found by WMAP for
\lcdm. The effect of
dynamical dark energy on $\sigma_8$
and on the formation of nonlinear structure offer promising routes
for further constraints on the time evolution of dark energy
or a possible falsification of the \lcdm\ model.

\begin{acknowledgments}
We acknowledge the use of the Legacy Archive for Microwave Background
Data Analysis (LAMBDA). Support for LAMBDA is provided by the NASA
Office of Space Science.
\end{acknowledgments}



\begin{thebibliography}{}
\bibitem{Astier:2005qq}
  P.~Astier {\it et al.},
  arXiv:astro-ph/0510447.

\bibitem{Riess:2004nr}
A.~G.~Riess {\it et al.}  [Supernova Search Team Collaboration],
Astrophys.\ J.\  {\bf 607}, 665 (2004)
[arXiv:astro-ph/0402512].

\bibitem{Percival:2001hw}
  W.~J.~Percival {\it et al.}  [The 2dFGRS Collaboration],
  Mon.\ Not.\ Roy.\ Astron.\ Soc.\  {\bf 327}, 1297 (2001)
  [arXiv:astro-ph/0105252].

\bibitem{Tegmark:2003ud}
M.~Tegmark {\it et al.}  [SDSS Collaboration],
Phys.\ Rev.\ D {\bf 69} (2004) 103501
[arXiv:astro-ph/0310723].

\bibitem{Spergel:2006hy}
  D.~N.~Spergel {\it et al.},
  arXiv:astro-ph/0603449.

\bibitem{Readhead:2004gy}
A.~C.~S.~Readhead {\it et al.},
Astrophys.\ J.\  {\bf 609} (2004) 498
[arXiv:astro-ph/0402359].

\bibitem{Goldstein:2002gf}
J.~H.~Goldstein {\it et al.},
Astrophys.\ J.\  {\bf 599}, 773 (2003)
[arXiv:astro-ph/0212517].

\bibitem{Dvali:2000hr}
  G.~R.~Dvali, G.~Gabadadze and M.~Porrati,
  Phys.\ Lett.\ B {\bf 485}, 208 (2000)
  [arXiv:hep-th/0005016].

\bibitem{Bekenstein:2004ne}
  J.~D.~Bekenstein,
  Phys.\ Rev.\ D {\bf 70}, 083509 (2004)
  [Erratum-ibid.\ D {\bf 71}, 069901 (2005)]
  [arXiv:astro-ph/0403694].

\bibitem{Wetterich:1994bg}
  C.~Wetterich,
  Astron.\ Astrophys.\  {\bf 301}, 321 (1995)
  [arXiv:hep-th/9408025].



\bibitem{Amendola:1999er}
  L.~Amendola,
  Phys.\ Rev.\ D {\bf 62}, 043511 (2000)
  [arXiv:astro-ph/9908023].

\bibitem{Amendola:2006qi}
  L.~Amendola, M.~Quartin, S.~Tsujikawa and I.~Waga,
  arXiv:astro-ph/0605488.

\bibitem{Doran:2002bc}
  M.~Doran and J.~Jaeckel,
  Phys.\ Rev.\ D {\bf 66}, 043519 (2002)
  [arXiv:astro-ph/0203018].

\bibitem{Wetterich:fm}
C.~Wetterich,
Nucl.\ Phys.\ B {\bf{302}}, 668  (1988)

\bibitem{Ratra:1987rm}
B.~Ratra and P.~J.~Peebles,
Phys.\ Rev.\ D {\bf{37}}, 3406  (1988)

\bibitem{Caldwell:1997ii}
R.~R.~Caldwell,~R.~Dave and P.~J.~Steinhardt,
Phys.\ Rev.\ Lett.\  {\bf{80}}, 1582 (1998)

\bibitem{Sahlen:2006dn}
  M.~Sahlen, A.~R.~Liddle and D.~Parkinson,
  arXiv:astro-ph/0610812.

\bibitem{Huterer:2002hy}
  D.~Huterer and G.~Starkman,
  Phys.\ Rev.\ Lett.\  {\bf 90}, 031301 (2003)
  [arXiv:astro-ph/0207517].

\bibitem{Doran:2000jt}
M.~Doran, M.~J.~Lilley, J.~Schwindt and C.~Wetterich,
Astrophys.\ J.\  {\bf 559}, 501 (2001)
[arXiv:astro-ph/0012139].

\bibitem{Doran:2001rw}
M.~Doran, J.~M.~Schwindt and C.~Wetterich,
Phys.\ Rev.\ D {\bf 64}, 123520 (2001)
[arXiv:astro-ph/0107525].

\bibitem{Bartelmann:2005fc}
  M.~Bartelmann, M.~Doran and C.~Wetterich,
  arXiv:astro-ph/0507257.

\bibitem{Caldwell:2003vp}
R.~R.~Caldwell, M.~Doran, C.~M.~Mueller, G.~Schaefer and C.~Wetterich,
Astrophys.\ J.\  {\bf 591} (2003) L75
[arXiv:astro-ph/0302505].

\bibitem{Ferreira:1997hj}
  P.~G.~Ferreira and M.~Joyce,
  Phys.\ Rev.\ D {\bf 58}, 023503 (1998)
  [arXiv:astro-ph/9711102].

\bibitem{Hebecker:2000zb}
  A.~Hebecker and C.~Wetterich,
  Phys.\ Lett.\ B {\bf 497}, 281 (2001)
  [arXiv:hep-ph/0008205].

\bibitem{Wetterich:2004pv}
C.~Wetterich,
Phys.\ Lett.\ B {\bf 594}, 17 (2004)
[arXiv:astro-ph/0403289].

\bibitem{Doran:2006kp}
  M.~Doran and G.~Robbers,
  JCAP {\bf 0606}, 026 (2006)
  [arXiv:astro-ph/0601544].

\bibitem{Bassett:2002qu}
  B.~A.~Bassett, M.~Kunz, J.~Silk and C.~Ungarelli,
  Mon.\ Not.\ Roy.\ Astron.\ Soc.\  {\bf 336}, 1217 (2002)
  [arXiv:astro-ph/0203383].

\bibitem{Corasaniti:2002vg}
P.~S.~Corasaniti and E.~J.~Copeland,
Phys.\ Rev.\ D {\bf 67}, 063521 (2003)
[arXiv:astro-ph/0205544].

\bibitem{Chevallier:2000qy}
  M.~Chevallier and D.~Polarski,
  Int.\ J.\ Mod.\ Phys.\ D {\bf 10}, 213 (2001)
  [arXiv:gr-qc/0009008].

\bibitem{Linder:2002et}
E.~V.~Linder,
Phys.\ Rev.\ Lett.\  {\bf 90}, 091301 (2003).



\bibitem{Vikman:2004dc}
  A.~Vikman,
  Phys.\ Rev.\ D {\bf 71}, 023515 (2005)
  [arXiv:astro-ph/0407107].

\bibitem{Hu:2004kh}
  W.~Hu,
  Phys.\ Rev.\ D {\bf 71} (2005) 047301
  [arXiv:astro-ph/0410680].

\bibitem{Huey:2004jz}
  G.~Huey,
  arXiv:astro-ph/0411102.

\bibitem{Caldwell:2005ai}
  R.~R.~Caldwell and M.~Doran,
  Phys.\ Rev.\ D {\bf 72} (2005) 043527
  [arXiv:astro-ph/0501104].

\bibitem{Zhao:2005vj}
  G.~B.~Zhao, J.~Q.~Xia, M.~Li, B.~Feng and X.~Zhang,
  Phys.\ Rev.\ D {\bf 72}, 123515 (2005)
  [arXiv:astro-ph/0507482].

\bibitem{MacTavish:2005yk}
  C.~J.~MacTavish {\it et al.},
  arXiv:astro-ph/0507503.


\bibitem{Dickinson:2004yr}
  C.~Dickinson {\it et al.},
  Mon.\ Not.\ Roy.\ Astron.\ Soc.\  {\bf 353} (2004) 732
  [arXiv:astro-ph/0402498].

\bibitem{Kuo:2002ua}
  C.~l.~Kuo {\it et al.}  [ACBAR collaboration],
  Astrophys.\ J.\  {\bf 600}, 32 (2004)
  [arXiv:astro-ph/0212289].

\bibitem{Eisenstein:2005su}
  D.~J.~Eisenstein {\it et al.},
  Astrophys.\ J.\  {\bf 633}, 560 (2005)
  [arXiv:astro-ph/0501171].

\bibitem{Doran:2006xp}
  M.~Doran, S.~Stern and E.~Thommes,
  arXiv:astro-ph/0609075.
\end{thebibliography}
\end{document}